\documentclass{article}

\usepackage[english]{babel}
\usepackage{multirow}
\usepackage[T1]{fontenc}

\usepackage[letterpaper,top=2cm,bottom=2cm,left=3cm,right=3cm,marginparwidth=1.75cm]{geometry}

\usepackage{amsmath}
\usepackage{graphicx}
\usepackage[colorlinks=true, allcolors=blue]{hyperref}
\usepackage{comment}
\usepackage{authblk}
\usepackage{textcomp}
\usepackage{float}
\usepackage[numbers, sort]{natbib}

\title{Deep Learning based Prediction of MSI using MMR Markers in Colorectal Cancer}

\author[1]{Ruqayya Awan}
\author[1,2]{Mohammed Nimir}
\author[1]{Shan E Ahmed Raza}
\author[1]{Mohsin Bilal}
\author[3]{Johannes Lotz}
\author[2]{David Snead}
\author[2]{Andrew Robinson}
\author[1,2,4]{Nasir Rajpoot}
\affil[1]{Department of Computer Science, University of Warwick, UK}
\affil[2]{Department of Pathology, University Hospitals Coventry \& Warwickshire, UK}
\affil[3]{Fraunhofer Institute for Digital Medicine MEVIS, L\"{u}beck, Germany}
\affil[4]{The Alan Turing Institute, London, UK}

\begin{document}
\maketitle

\begin{abstract}

The accurate diagnosis and molecular profiling of colorectal cancers are critical for planning the best treatment options for patients. Microsatellite instability (MSI) or mismatch repair (MMR) status plays a vital role in appropriate treatment selection, has prognostic implications and is used to investigate the possibility of patients having underlying genetic disorders (Lynch syndrome). NICE recommends that all CRC patients should be offered MMR/MSI testing. Immunohistochemistry is commonly used to assess MMR status with subsequent molecular testing performed as required. This incurs significant extra costs and requires additional resources. The introduction of automated methods that can predict MSI or MMR status from a target image could substantially reduce the cost associated with MMR testing. Unlike previous studies on MSI prediction involving training a CNN using coarse labels (Microsatellite Instable vs Microsatellite Stable), we have utilised fine-grain MMR labels for training purposes. In this paper, we present our work on predicting MSI status in a two-stage process using a single target slide either stained with CK8/18 or H\&E. First, we trained a multi-headed convolutional neural network model where each head was responsible for predicting one of the MMR protein expressions. To this end, we performed the registration of MMR stained slides to the target slide as a pre-processing step. In the second stage, statistical features computed from the MMR prediction maps were used for the final MSI prediction. Our results demonstrated that MSI classification can be improved by incorporating fine-grained MMR labels in comparison to the previous approaches in which only coarse labels were utilised.

\end{abstract}

\section{Introduction}

The molecular profiling of colorectal cancers (CRC) plays an increasingly important role in determining the most appropriate treatment and management options for patients. In particular, microsatellite instability (MSI) and mismatch repair (MMR) deficiency has been identified as an important factor which has prognostic implications and can impact treatment options. The National Institute for Clinical Excellence (NICE) currently recommends that all CRC patients are offered testing using immunohistochemistry (IHC) or MSI testing. 

The presence of MMR deficiency may infer a more favourable prognosis when compared to patients with stage adjusted MMR proficient colorectal adenocarcinomas (CRA). Studies have also shown that patients with early stage MMR deficient CRA may not benefit from adjuvant chemotherapy with 5-fluorouracil.  Patients with MMR deficient cancers may also be eligible for checkpoint blockade therapy (immunotherapy) in certain circumstances \cite{lee2016mismatch, luchini2019esmo}. Immunotherapy has been shown to be a compelling treatment for some cancers when other traditional treatments such as chemotherapy become less effective. However, the number of patients who  would benefit from immunotherpay is limited and there are predictive biomarkers that can be used for the selection of these patients. Its deficiency in the tumour is correlated with the efficacy of immunotherapy due to a high level of mutations [3]. This correlation has been extensively studied in CRC and similar findings are observed in other tumour types as well. MMR/MSI testing also plays an important role in identifying patients who may suffer from underlying genetic disorders such as Lynch Syndrome which may have significant implications for their relatives.

The MMR system is a sophisticated system for maintaining the integrity of DNA during replication and recombination and hence is responsible for genomic stability [4]. It comprises MMR proteins which are dependent on four genes: mutL homologue 1 (MLH1), mutS homologue 2 (MSH2), mutS homologue 6 (MSH6) and postmeiotic segregation increased 2 (PMS2). These proteins repair the DNA by identifying errors occurring at microsatellites which are short repetitive DNA sequences. The differences in the length of the microsatellite between normal and tumour tissues of the same patient would indicate the deficiency in MMR and would lead to mutations of the above-mentioned genes.   Tumours with variation in microsatellite length are known as MSI tumours and those with a high level of variations are termed MSI high (MSI-H). There are two main methods for assessing the MMR status of the tumour: 1) IHC testing which is mainly used in clinical settings to detect MMR proteins and 2) polymerase chain reaction (PCR) based methods for testing MSI. For PCR testing, the National Cancer Institute (NCI) workshop in 1997 recommended following the Bethesda guidelines [5] on the five microsatellite markers known as the Bethesda markers. These markers consist of three dinucleotide loci (D17S250, D2S123 and D5S346) and two mononucleotide loci (Big Adenine Tract [BAT]-25 and BAT-26). Tumours showing instability at two or more markers are considered as MSI-H and when only one of the Besthesda markers is unstable then the tumour is interpreted as MSI-L. If none of the markers is unstable then the tumour is considered microsatellite stable (MSS). Both IHC and PCR testing methods may have false-negative rates and are hence complementary to each other. However, IHC testing gives us additional information in the form of stained slides that can be utilized to see if it improves the performance of the computational method for MSI prediction.

MMR expression status is determined by immunohistochemistry (IHC) for MMR proteins. The IHC results are formed by assessing whether there is an expression or complete loss of staining within tumour nuclei rather than scoring the intensity of staining as in ER/PR or HER2 IHC testing. If there is a loss of expression for MLH1 or MSH2 proteins, this is often accompanied by loss of expression of their respective partners PMS2 or MSH6. However, isolated loss of expression of PMS2 or MSH6 will not affect their partners since they can bind to other molecules to maintain their stability [6]. DAB (3,3\textquotesingle -diaminobenzidine) immunohistochemistry produces a dark brown product that can be linked to the antibody being tested. A pathologist would assign a loss or intact status to each MMR stained slide based on the presence or absence of DAB stain. Identifying all four proteins as intact indicates proficient MMR (pMMR) status while loss of expression in any one of the proteins shows that the patient may have a germline mutation causing MMR deficiency and hence indicates deficient MMR (dMMR) status.

Our study was focused on the detection of CRC with dMMR status.  We identified these cases using tissue slides stained with haemotoxylin and eosin (H\&E) and cytokeratin 8 (CK8) paired with cytokeratin 18 (CK18), referred to as CK8/18. In previous studies, computer neural networks (CNN) were trained using MSI/MSS labels whilst we have utilised MMR labels for training purposes in a two-stage process. On providing fine- grain MMR labelling, we have shown that our approach improves the performance of MSI/MSS classification in comparison to previous approaches.  


\section{Literature Review}

The current methods for MMR testing of CRC incur additional costs and use of resources for pathology departments. Recently, researchers have been able to predict the MSI status directly from the H\&E stained slides using deep learning and their encouraging results demonstrate that such an automatic system could enable MSI screening for every patient worldwide and potentially reduce some of these additional costs. Kather \textit{et al.} \cite{kather2019deep} employed a trained CNN to classify tumour patches in a WSI into MSI or MSS class and finally, the class with a higher number of patches was used for assigning the WSI label. They later refined and tested this system on a large-scale data \cite{kather2020pan, echle2020clinical, muti2021development, echle2022artificial}, they perform MSI detection using the same approach without performing tumour segmentation as a preprocessing step.

Schmauch \textit{et al.} \cite{schmauch2020deep} proposed a deep learning-based pipeline for predicting RNA-Seq profiles from H\&E stained WSIs. The proposed pipeline comprises two main steps. First, they perform CNN based feature extraction for selected 8000, randomly selected, tissue tiles, resulting in a feature map of size 8000$\times$2048 for each WSI. In the second step, they train a multi-layer perceptron referred to as HE2RNA for predicting gene expression for each tile. During the training stage, they take an average of different numbers (\textit{k}) of the highest tile predictions at every iteration for deciding slide-level labels. At the inference stage, the slide-level label is obtained by taking an average of all tile-level labels obtained with every possible value of \textit{k}. For MSI prediction, they simplified the above approach by averaging CNN features for every tile resulting in a feature map of size 1$\times$2048 for every WSI and also modified the HE2RNA model accordingly.

Bilal \textit{et al.} \cite{bilal2021novel} proposed a novel deep learning framework for predicting the status for a number of molecular pathways, including MSI. Their proposed framework comprises three models: The first model (ResNet18) performs tumour segmentation, the second model referred to as iterative draw-and-rank sampling (IDaRS) takes random (\textit{r}=45) and top (\textit{k}=5) tiles from each training slide for fine-tuning a CNN model (ResNet34) for MSI prediction. A trained IDaRS model then selects the tiles with the most discriminatory features. The last model (HoVer-Net \cite{graham2019hover}) performs nuclei segmentation and classification of strongly predicted tiles into five different cell types. To estimate the differential feature importance of each cell type in the strongly predicted tiles of MSI (or other molecular pathways), a linear support vector machine (SVM) was trained. Cell count for each category was used as a feature set for training an SVM. Their proposed method achieved an AUROC value of 0.90 which is significantly better than that of the state-of-the-art method \cite{kather2019deep} by 0.13.

\section{MMR Panel Interpretation}
\begin{figure}[h]
\centering
\includegraphics[scale=0.75]{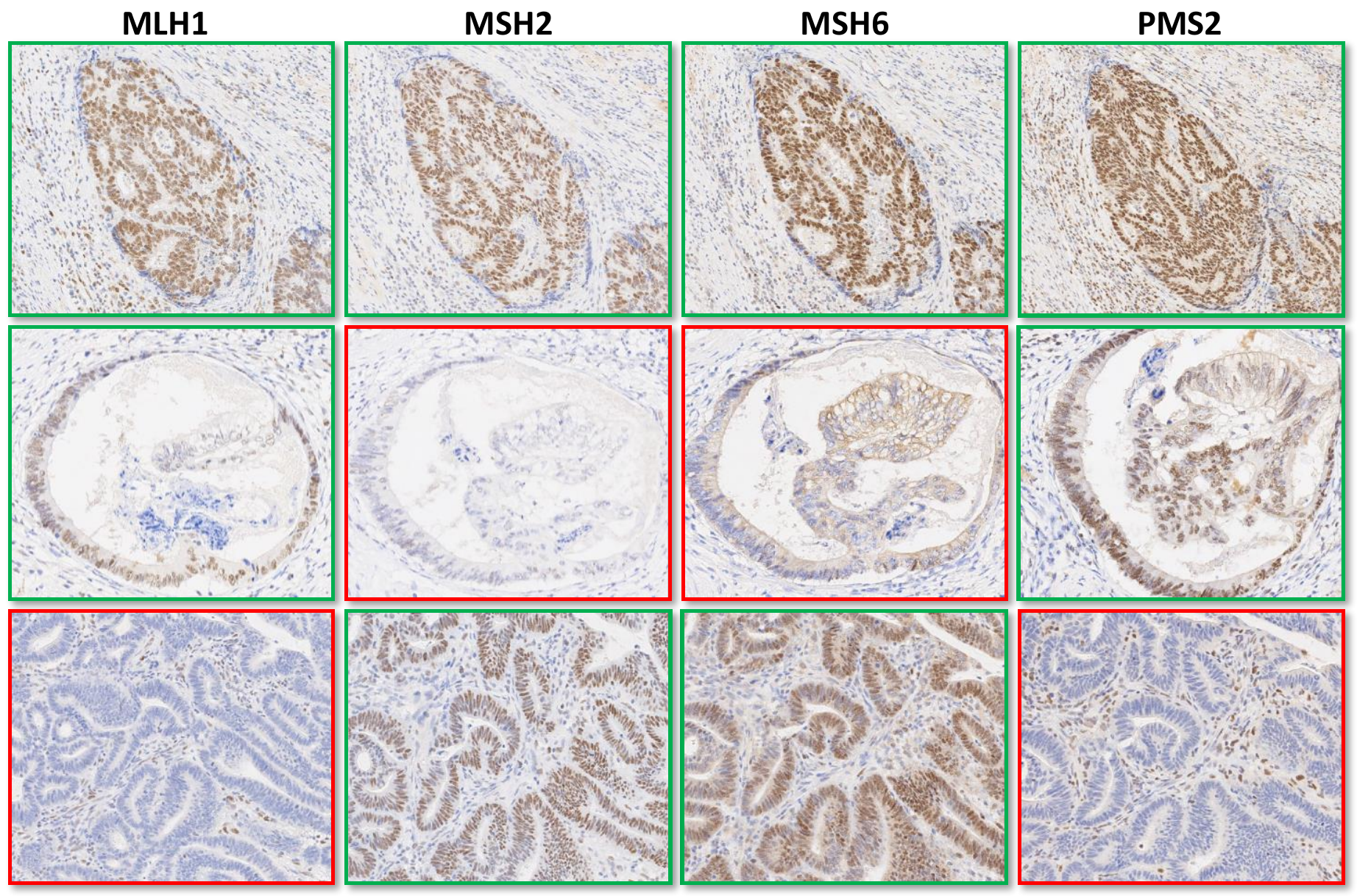}
\caption{Visual fields taken from three cases with different MMR staining patterns. The border colour represents the MMR status where green and red are used for patches extracted from WSIs with intact and loss status respectively. Top row: intact status for all MMR markers; Middle row: intact status for MLH1 and PMS2 while loss status for MSH2 and MSH6; Bottom row: loss of MLH1 and PMS2 while the intact status for MSH6 and PMS2. Notice the nuclear staining of internal positive controls (lymphocytes and stromal cells).   }
\label{fig:mmr_expression}
\end{figure}

The four MMR proteins exhibit nuclear staining patterns and are assessed for the presence and absence of brown DAB staining. MMR staining should be observed in normal cells such as lymphocytes, fibroblasts and normal epithelium. However, the MMR expression is only evaluated in tumour cell nuclei whilst comparing their staining to that of normal cells acting as internal controls. Internal controls validated for the presence of staining are termed internal positive controls. In the absence of staining in internal control, MMR protein expression cannot be evaluated and needs repeated IHC testing either on the same or a different tissue block. A clear loss of nuclear staining in tumour cells, punctate, or focal weak equivocal staining,  in the presence of valid internal positive controls, would indicate a loss of protein expression, whilst a sample with unambiguous staining of tumour nuclei at least equal to or more than that of valid internal positive controls would be identified as intact protein expression. In Figure \ref{fig:mmr_expression}, visual fields extracted from WSIs of various staining patterns are shown.

Typically, appropriately fixed MMR samples tend to show a strong DAB signal. However, variability in the intensity of DAB staining can be observed. Whilst MMR protein expression should be localized to the nucleus, there can sometimes be abnormal cytoplasmic staining patterns. The possible causes of such unusual occurrences could be issues in IHC testing or a recently reported germline abnormality resulting in fusion of EPCAM and MSH2 \cite{sekine2017cytoplasmic}. In the presence of both cytoplasmic and nuclear staining, the IHC test may be repeated on the same block or a different block whilst only a cytoplasmic staining pattern is considered a loss of expression.

In a sample with intact status, it is not uncommon to have a tissue area with the absence of protein staining. There are no standard criteria on an exact cutoff for normal protein expression. There are different propositions in the literature for a cutoff of 1\%, 5\% or 10\% \cite{pai2016practical} while The College of American Pathologists suggests that any amount of positive staining in the tumour nuclei is indicative of intact status \cite{burgart2021template}.

\section{The Proposed Method}
\begin{figure}[h]
\centering
\includegraphics[scale = 0.4]{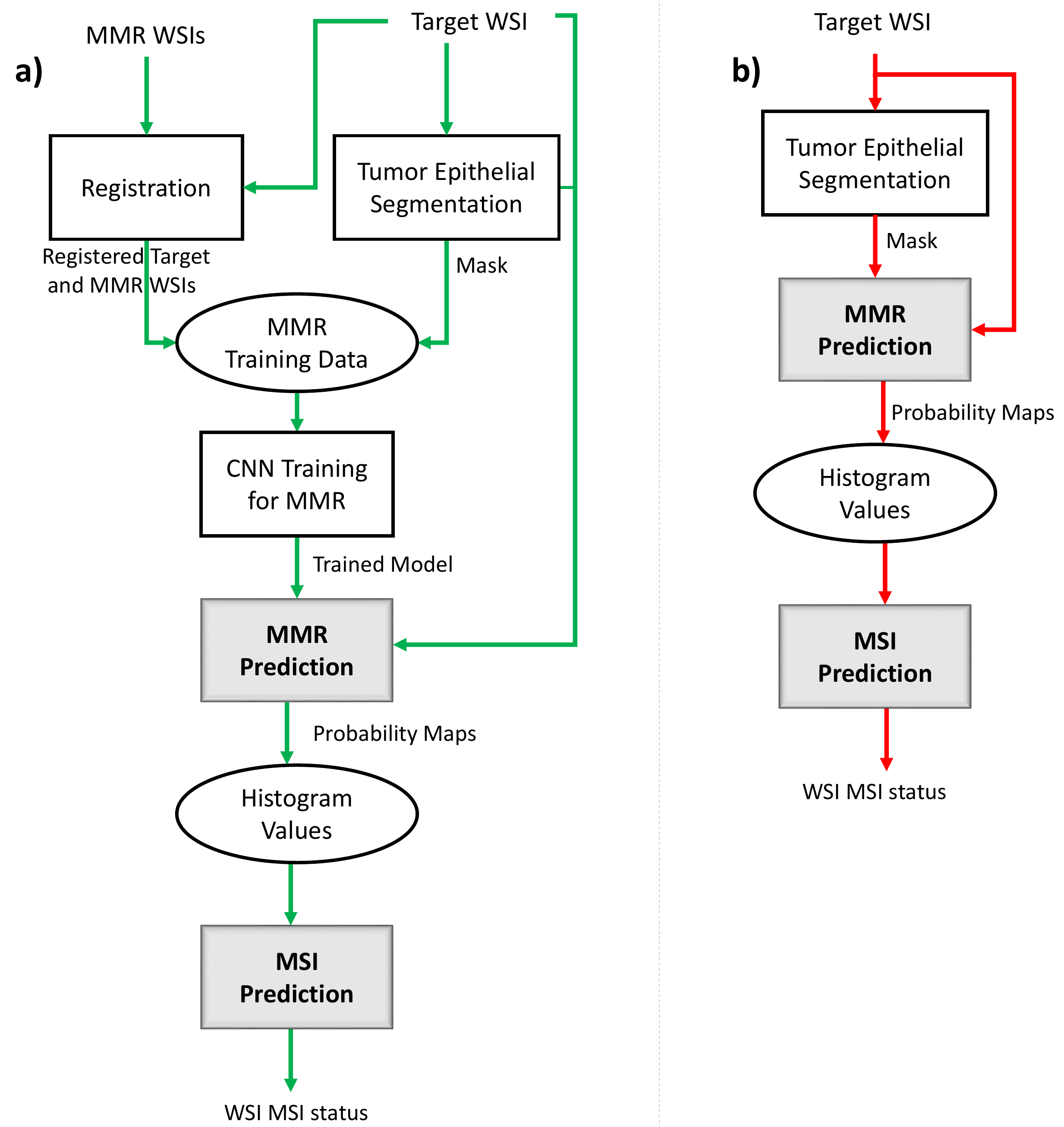}
\caption{A block diagram showing the overall flow for the a) training and b) testing of our 2-step MSI classification pipeline. In this study, we experimented with CK8/18 and H\&E as target WSIs. }
\label{fig:block_diagram}
\end{figure}

A block diagram showing the overall flow for training and inference of the proposed method for prediction of the MSI status is shown in Figure \ref{fig:block_diagram}. We performed a number of pre-processing steps: registration of MMR slides to the target image and tumour and epithelial segmentation. During registration, we spatially translated the H\&E and MMR slides to align with the CK8/18 slide which is chosen as a reference image in this work. We perform tumour epithelial region segmentation in order to restrict our MMR analysis to the epithelial area in cancerous tissue. This is because the pathologists analyse the nuclear staining of cancerous epithelial nuclei while deciding the status (intact/loss) of MMR slides. To this end, we perform tumour segmentation followed by epithelial segmentation. These two masks are then used to get epithelial regions in the tumour area.

\subsection{Registration}
\label{reg_section}
A registration step in a co-localisation analysis workflow should be able to allow a significant spatial overlap between the two images such that the location of corresponding tissue structures can be determined. To this end, we employed a pipeline proposed in \cite{awan2022deep}. The pipeline comprised of three main steps: pre-processing, estimation of rigid alignment using `Deep Feature based Registration' (DFBR) method, followed by a non-linear registration. During the pre-processing step, we generated a tissue mask for an image pair and modified the input images such that they appear spatially similar. Our DFBR method further contains three sub-modules: \textit{pre-alignment} to perform rough alignment, \textit{tissue transform} estimates the transformation parameters using cropped tissue region and \textit{block-wise transform} refines \textit{tissue transform} by performing feature matching in a block-wise manner. After an image is registered using our DFBR, we observed a slight offset in some cases. To fix this offset, we added a local transform module which is followed by an existing non-linear registration method \cite{lotz2021high}. 



\subsection{Tumour Segmentation}
We trained a classification model using a publicly available colorectal adenocarcinoma dataset released with \cite{shaban2020context}. This dataset consists of 300 visual fields extracted at 20$\times$ magnification of size at 7000$\times$5000 pixels. These visual fields either belong to normal tissue or cancerous tissue. The cancerous visual fields were further classified into two categories: low-grade and high-grade cancer. Of the 300 images, normal and low-grade classes comprised 120 images per class while the high-grade category comprised 60 images. We extracted patches of size 224$\times$224 pixels for network training using a tumour mask. Tumour masks for the CK8/18 images were generated by applying geometrical transformation on the H\&E masks. We used transformation parameters which were computed for registering the H\&E image to the CK8/18 image, as discussed in section \ref{reg_section}. A transformed H\&E mask overlaid on CK8/18 is shown in Supplementary Figure 1. An Xception model was trained using an Adam optimiser with cross-entropy as an objective function. 



\subsection{Epithelial Segmentation}
The MMR analysis should be restricted to the epithelial areas in cancerous tissue, since the presence or absence of positive staining in cancerous epithelial nuclei determines the status (intact/loss) of MMR markers. Taking this into consideration, one possible approach could be to perform epithelial nuclei segmentation. To this end, we used a publicly available dataset known as CoNSeP for training a cell segmentation network known as HoVerNet \cite{graham2019hover}. This dataset consisted of 41 visual fields taken from 16 H\&E stained colorectal adenocarcinoma WSIs. Each visual field was of size 1000$\times$1000, extracted at 40$\times$ magnification. We performed stain deconvolution on input images to extract the H channel so that the same network could be used for the H\&E and CK8/18 stained images. However, we didn't use nuclei segmentation masks for our analysis since the trained model underperformed. An example image from a CK8/18 slide, its H channel and Hovernet predicted nuclei segmentation mask overlaid on H channel are shown in Supplementary Figure 2. Another possible approach could be segmenting the epithelial layer which is less complicated in comparison to nuclei segmentation in terms of computational time and labelled data creation for training a model. 


Unlike tumour segmentation, we didn't transform the H\&E mask for generating the CK8/18 mask. Instead, we followed different approaches for CK8/18 and H\&E images. This is due to the differences in tissue architecture (mostly visible at higher magnification levels) between H\&E and CK8/18 images. We performed epithelial segmentation for H\&E images using a UNET-like architecture and used a publicly available dataset known as CRAG \cite{awan2017glandular, graham2019mild}. This dataset consisted of 213 visual fields of size 1500$\times$1500 pixels at 20$\times$ magnification, provided in two splits for training and validation. The train and valid sets comprised 173 and 40 visual fields, respectively. We manually annotated the epithelial layer in these images. To improve the model performance on the COMET dataset, we also added 23 visual fields of size 2048$\times$2048 pixels from the COMET dataset for training purposes. We performed our experiments using input images at 20$\times$ and 10$\times$ magnification and didn't find a significant difference between their average DICE values. To accelerate the mask generation step, we finally used a model trained with images obtained at 10$\times$ magnification. We obtained a mean DICE value of 0.8645 on the CRAG validation set. The normal epithelial region was excluded by multiplying the predicted mask with the tumour mask.

Generating an epithelial mask for CK8/18 images was relatively straightforward. CK8/18 are known to have a cytoplasmic and membranous expression in glandular epithelial cells (both normal and cancerous). Considering this, an initial mask was created by a thresholding operation on the DAB channel and then we multiplied this mask with the tumour mask to get a final mask for the tumour epithelial region. The predicted epithelial masks overlaid on the original CK8/18 and H\&E stained images are shown in Supplementary Figure 3. 


\subsection{MSI/MSS Classification}
In contrast to previous studies on MSI prediction, our proposed approach makes use of the additional information (MMR labels) for predicting MSI status. We performed MSI prediction in two steps: MMR prediction followed by MSI prediction.

\subsubsection{MMR Prediction}
There could be at least two different strategies for learning MMR labels, either train separate CNN models for predicting each MMR label individually (as in \cite{gamble2021determining} for ER/PR/HER2 prediction) or train a single multi-output CNN model. In this work, we opted for the latter approach by implementing a multi-headed CNN architecture where each head is responsible for predicting a particular MMR staining expression status. The proposed CNN architecture is shown in Supplementary Figure 4. We employed a partial Xception network as a backbone model while removing the last fully connected layer. The default input size for the Xception model is 299$\times$299 pixels. However, we selected a patch size of 512$\times$512 pixels to include a bit more context. With an input size of 299$\times$299 pixels, the feature map generated by the last layer before pooling is of dimension 10$\times$10$\times$2048 while with an input of 512$\times$512 pixels, a feature map of size 16$\times$16$\times$2048 pixels is generated. To reduce the spatial dimension of the resulting feature map, we added a convolutional layer before the global average pooling layer. After global pooling, the network is divided into four heads, each head comprises three fully connected layers. To prevent the model from overfitting, we added a dropout layer in the first two dense layers with a dropout rate of 0.5.

\begin{figure}
\centering
\includegraphics[width=\linewidth]{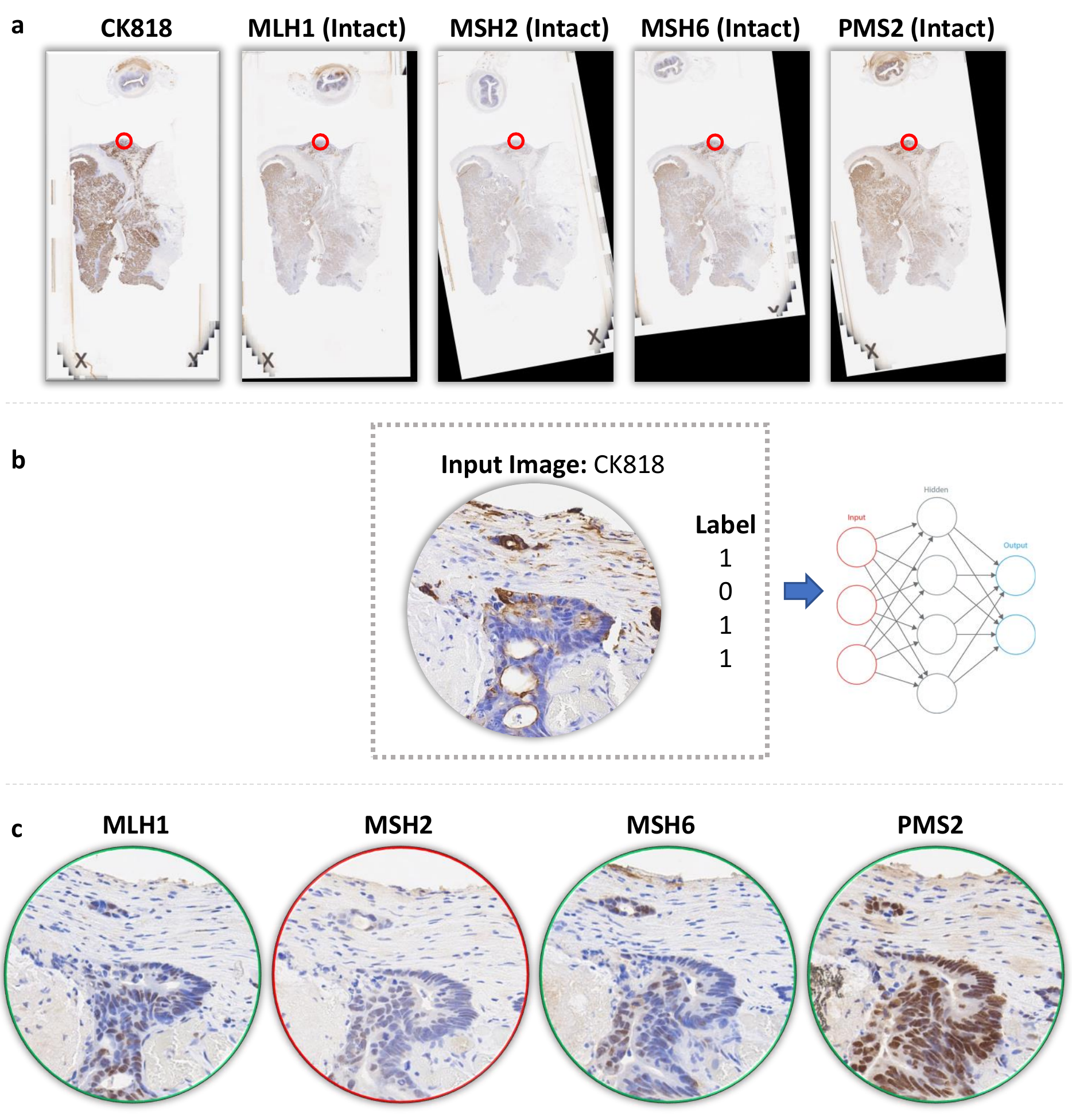}
\caption{Co-localised MMR Prediction using CK8/18 as a target image. a) Downsampled WSIs of MMR IHC sections registered to CK8/18 stained WSI. b) An example input image used for training a CNN. Its ground truth label is defined by thresholding Hscore values computed for the corresponding patches from MMR slides. The corresponding MMR patches are shown in c). The colour of the borderline in c) shows the expression status for the given patch. Green represents that the patch is positively stained with DAB while red shows the absence of DAB stain. Patches shown in b) and c) are highlighted in red circles in a). }
\label{fig:patch_labelling}
\end{figure}

\label{mmr_training}
The ground truth data for the COMET dataset comprises MMR expression status (intact/loss) for a WSI. If we follow a weakly supervised strategy for training, then the WSI label could be assigned to each patch extracted from that particular WSI. However, as mentioned earlier, the MMR stained tissue slides with intact status are likely to have some tumour regions without staining or variable staining. Taking this into consideration, we adopted stain co-localisation for generating labels locally, rather than globally. We term this as a `localised' setting for generating training data. Using this setting, we would get four binary labels against each input patch, where each label corresponds to an individual MMR protein. To this end, we first registered the MMR slides to the target slide (CK8/18 or H\&E). We then extracted a patch from a target slide at 20$\times$ magnification using a tumour epithelial mask and the corresponding patches from the registered MMR slides. If the MMR expression label for a WSI was `Loss' then we assigned the `Loss' label to every patch of that WSI. However, if the MMR expression label was Intact then we first computed the DAB stain intensity of a patch using the H-Score \cite{mcclelland1990automated} and then performed thresholding to decide if the patch had loss or intact status. A pictorial demonstration of generating labelled data is provided in Figure \ref{fig:patch_labelling}. To compute the H-score, cellular staining intensity is first divided into four categories: strong (3+), moderate(2+), weak(1+) and no staining/negative(0). The summation of the weighted percentage of these intensity categories would compute the final H-Score value. We performed the following steps for computing the H-Score: 

\begin{enumerate}
  \item Deconvolve the IHC patch into H and DAB channels;
  \item Threshold H channel to detect nuclei; 
  \item Using nuclei mask, we applied thresholding to each nuclei pixel in the DAB channel such that they fall into one of the four categories: strong ($\geq$0 \& $\leq$60), moderate ($\geq$61 \& $\leq$150), weak ($\geq$151 \& $\leq$200) and negative ($\geq$201 \& $\leq$255). Thresholding is performed after taking the complement of the DAB channel;
  \item Finally, we computed the H-Score using the following formula \cite{detre1995quickscore, liu2018end}:
   \begin{center}
        H-Score = 3$\times$(\% of nuclei stained at intensity category `3+') +\\ 
        \qquad \qquad \quad 2$\times$(\% of nuclei stained at intensity category `2+') +\\
        \qquad \qquad \quad 1$\times$(\% of nuclei stained at intensity category `1+') +\\
    \end{center}
  \item The H-Score ranges between 0 and 300. On the basis of a discriminatory threshold, the patch status was considered either intact or loss. We selected a threshold value of 130.
\end{enumerate}

We then trained our multi-headed network using target input patches generated using our localized  setting. We trained the model using Adam optimiser with binary cross-entropy as an objective function for each MMR label. We performed stain augmentation, rotation, flipping and elastic distortion for augmenting our training patches. Once the network was trained, we proceeded to our second step for MSI prediction where we employed the trained model for generating MMR probability maps. We then extracted features from these maps and utilised them for classifying each case as MSI or MSS.
 
\subsubsection{MSI Prediction}
\label{MSI_prediction}

\begin{figure}[h]
\centering
\includegraphics[width=\linewidth]{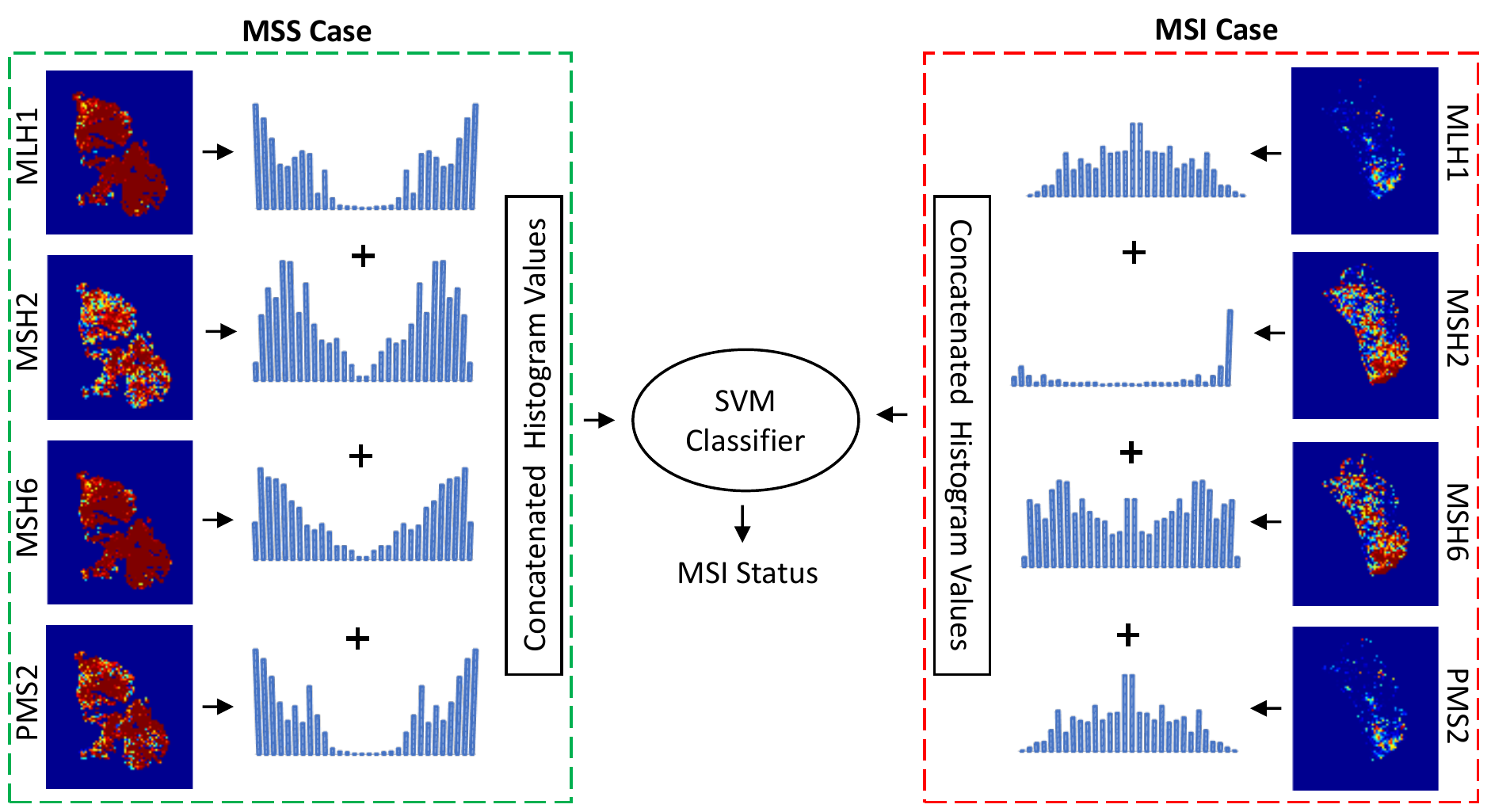}
\caption{Overview of the proposed MSI prediction step. The MMR probability heatmaps and corresponding histogram of probabilities are shown for both MSS and MSI cases.}
\label{fig:msi_prediction}
\end{figure}

Using the trained MMR model, we generated probability maps for all the train and test cases. For any given case, we produced four probability maps, one per MMR biomarker. We then computed features for each MMR protein which is the histogram of probabilities of tumour epithelial patches. We used two histograms: one for protein intact status and the other for protein loss status. We selected the number of histogram bins parameter as 15 and normalised the histogram values. Finally, we inputted these features to an SVM classifier for WSI level MSI prediction. A diagrammatic illustration of MSI prediction step is shown in Figure \ref{fig:msi_prediction}. Note that MMR slides are only used during the training phase for labelled data generation. During inference we only used target slides for MMR and MSI prediction, hence ceasing the time burden of the pre-processing step of registration.

\section{Dataset and Performance Measures}
We used the COMET dataset obtained from the University Hospitals Coventry and Warwickshire (UHCW) NHS Trust in Coventry, UK. The dataset was provided after de-identification and informed consent was obtained from the patients. It contained 86 cancerous cases (from different patients) which were used for our experiments on MSI prediction. There were 16 slides per case, each scanned using an Omnyx VL120 scanner at 0.275 microns/pixel. These slides were stained with different stains: CK8/18, Ki67, p53, Vimentin, MMR (MLH1, MSH2, MSH6 and PMS2), Ecadherin, EpCAM, PTEN and H\&E and their exact sequence of staining is shown in Supplementary Figure 5. We considered six slides per case, involving MMR prediction from H\&E and CK8/18 images for the end purpose of MSI prediction. For WSI-level MMR labels, we provided MMR tissue images along with their corresponding H\&E to a pathologist. Of the 72 tumours out of 86, in which MMR analysis was possible, we obtained the expression status for each MMR slide. We then categorised these 72 cases into MSI and MSS using their MMR labels. Cases for which all the MMR slides were given intact status were classified as MSS whilst any case with at least one MMR slide with loss status was categorised as MSI. In total, we obtained 16 (22.2\%) MSI and 56 (77.8\%) MSS cases. The count of cases with different MMR and MSI ground truth labels are presented in Supplementary Table 1. We assessed the performance of our method using 5-fold cross-validation while keeping the same train and test set for CNN (MMR prediction) and SVM (MSI prediction) training and testing. We evaluated the patch-level MMR prediction using DICE overlap and WSI-level MSI prediction using AUROC.



\section{Results and Discussion}
We carried out two sets of experiments: one with a CK8/18 slide as a target image and the other with a H\&E slide as a target image. Below, we present our results of MMR and MSI classification steps for both CK8/18 and H\&E slide images.

\subsection{MMR Prediction}
\begin{figure}
\centering
\includegraphics[width=\linewidth]{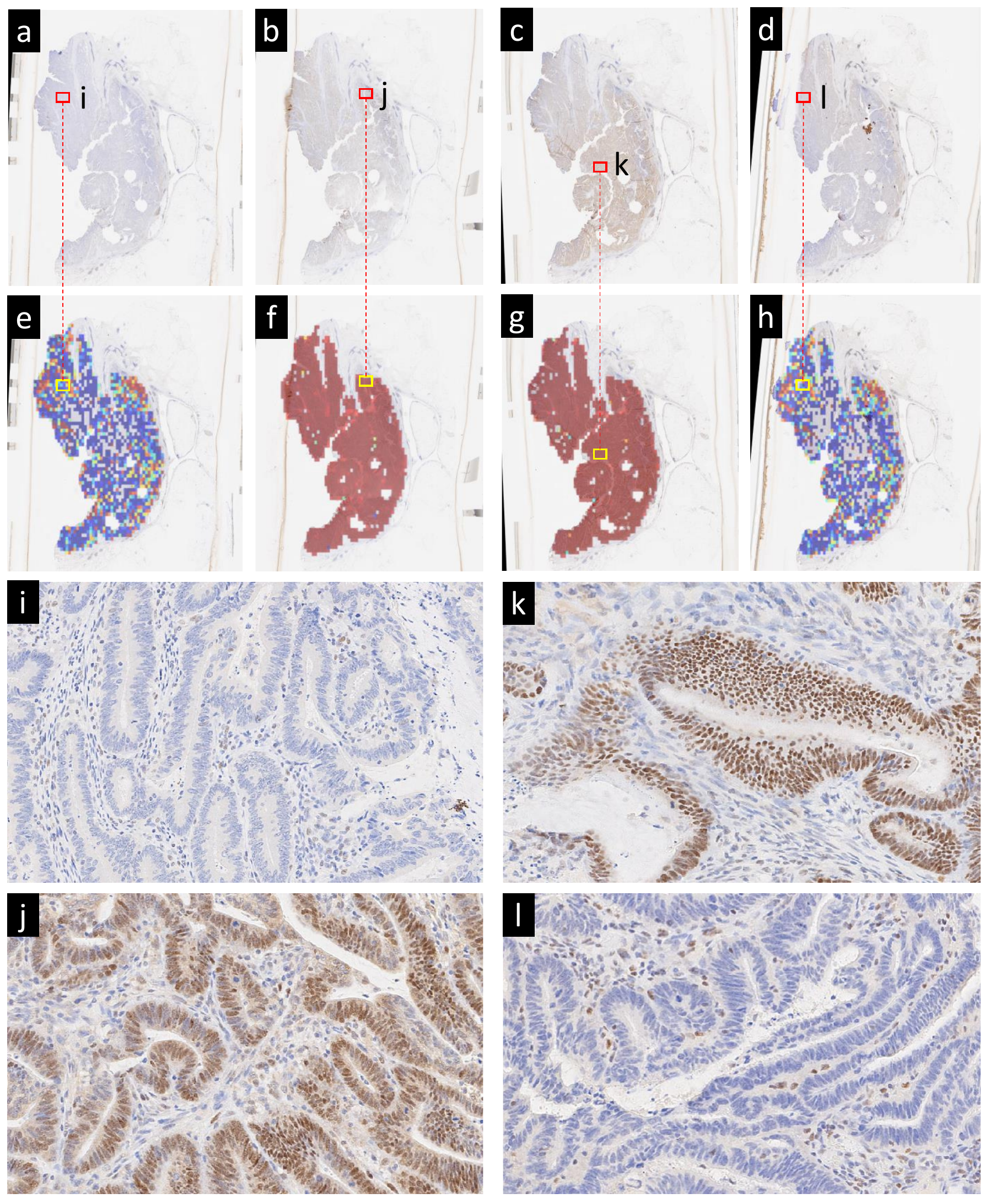}
\caption{MMR prediction against CK8/18. a-d) MMR stained slides: MLH1, MSH2, MSH6 and PMS2, e-f) respective probability heatmaps overlaid on registered MMR stained WSIs, i-l) zoomed in patches highlighted on a-h with small rectangles. The ground truth MMR expression status for a-d is Loss, Intact, Intact and Loss.}
\label{fig:mmr_pred_ck818}
\end{figure}

The MMR probability maps for intact status overlaid on the original MMR tissue images are shown in Figure \ref{fig:mmr_pred_ck818}, along with the zoomed-in regions. We first computed the DICE overlap ratio to evaluate the performance of our multi-headed CNN model for MMR prediction. To this end, we generated the ground truth H-Score maps by processing MMR slides at 20$\times$ magnification in a sliding window fashion. The non-epithelial regions were excluded while computing the H-Score map. Thresholding was then performed to convert the H-Score maps to binary maps where the foreground represents a region with intact status whilst the background represents a non-tissue region or a region with loss status. The network-generated MMR probability maps were also converted to binary maps using a threshold value of 0.5. 


\begin{figure}[h]
\centering
\includegraphics[width=\linewidth]{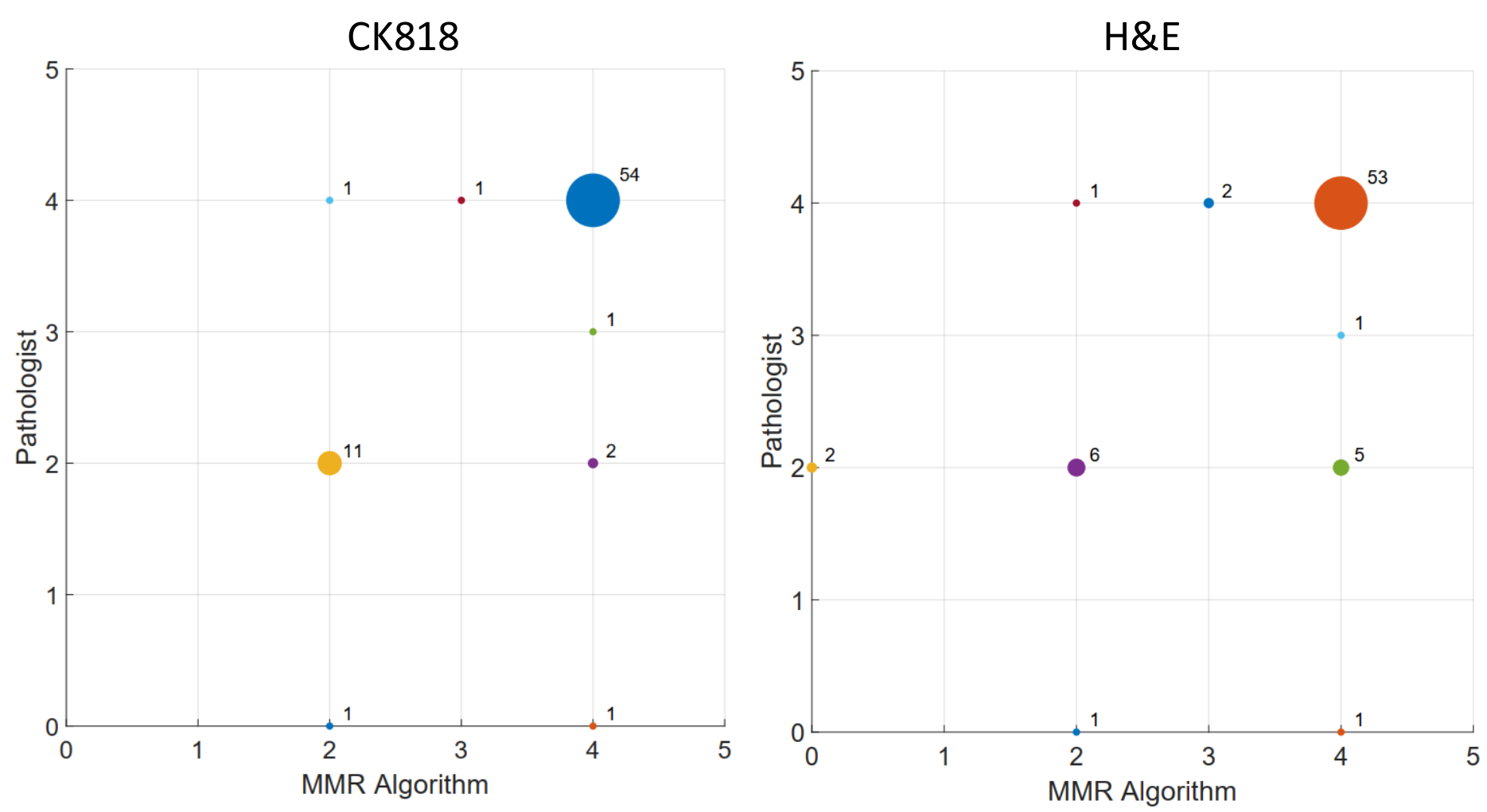}
\caption{Scatter plot showing the WSI-level MMR prediction results using our baseline experiment in comparison to that of the pathologist's assignment. Any MMR slide with intact status is given a value of 1 whilst a slide with loss status is given 0. This plot was generated by taking the sum of these MMR values. The size of the circular markers is relative to the number of cases with matching pathologist's and algorithm generated labels.}
\label{fig:scatter_plot_mmr}
\end{figure}

We also performed a baseline experiment by directly utilising the MMR prediction maps for MSI classification, without training an SVM model. Thresholded probability maps were used for deciding intact/loss status for MMR proteins and it was based on the number of patches with intact/loss status. The intact status was assigned if the number of intact patches was more than the number of patches with loss status. Finally, the MSI status was identified using these MMR labels. We obtained AUC of 0.8869 ± 0.150 and 0.8500 ± 0.124 for CK8/18 and H\&E respectively, across 5-fold cross-validation data split. Figure \ref{fig:scatter_plot_mmr} shows the results of MMR status prediction and its correlation with the pathologists' MMR labels. On using MMR labels for CK8/18 as a target image, we were able to identify 98\% and 75\% of MSS and MSI cases, respectively. Whilst for H\&E as a target image, we correctly classified 95\% of MSS and 56\% of MSI cases, respectively.
 
\subsection{MSI Prediction}
\begin{figure}[h]
\centering
\includegraphics[width=\linewidth]{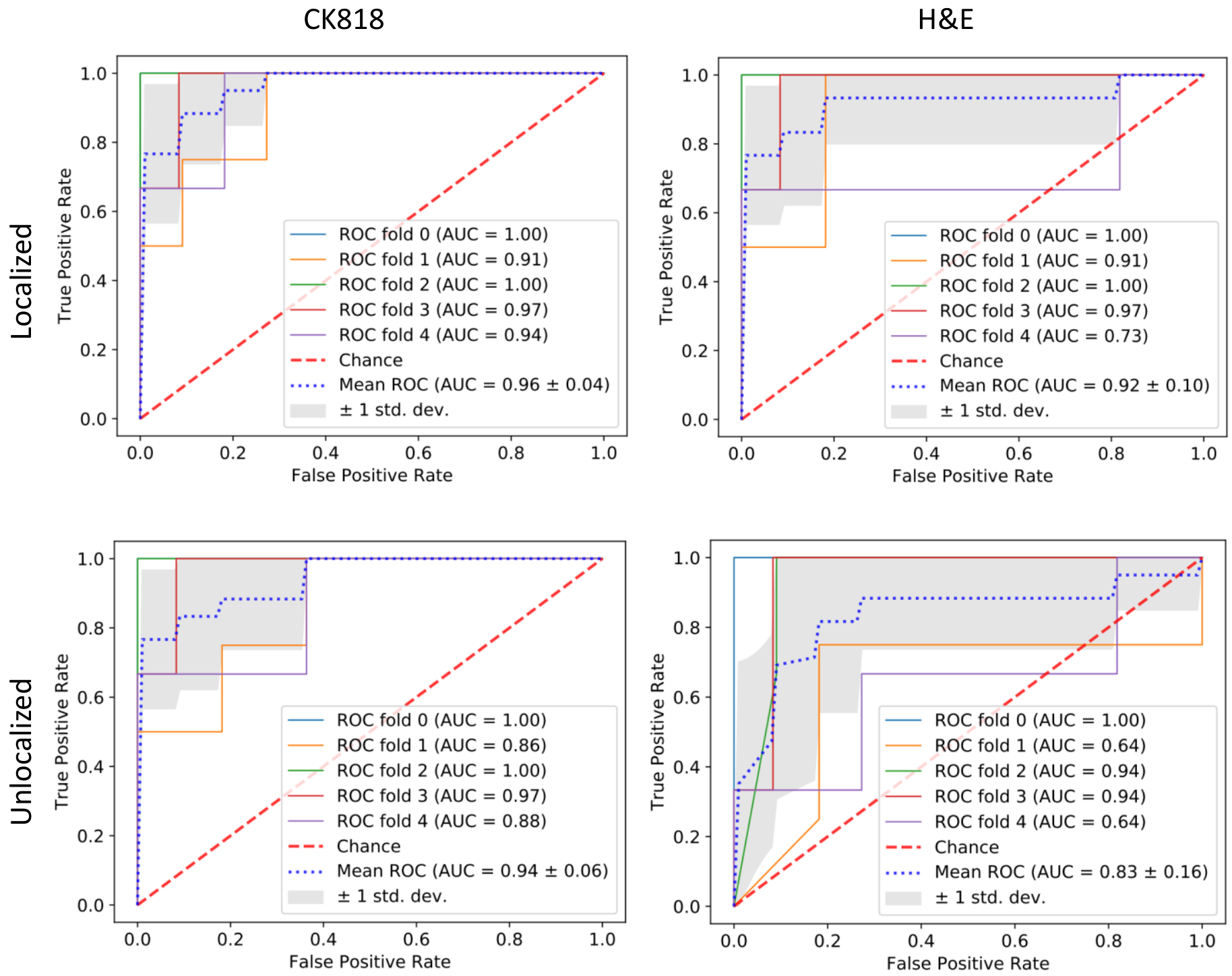}
\caption{MSI Prediction on CK8/18 and H\&E images (COMET dataset) using localised and unlocalised settings. AUROC plots of 5-fold cross-validation along with average AUROC is shown. For each curve, x- and y-axis represent false positive and true positive rates.}
\label{fig:roc_msi}
\end{figure}

We trained SVM models using histogram features with different kernel functions: linear, polynomial and RBF. We utilised our multi-headed network trained using two different settings for extracting histogram values. The localised setting is the one discussed in Section \ref{mmr_training} whilst for the unlocalised setting, slide-level MMR labels were assigned to all the patches rather than performing stain co-localisation analysis for generating labels. The average AUCs are reported in Table \ref{tab:auc_ck818_he}. The SVM with RBF kernel has been shown to outperform other kernel functions for both CK8/18 and H\&E irrespective of any training setting. Our results suggest that the histogram values generated in the localised setting are more effective in classifying cases into MSI/MSS status as compared to those generated in an unlocalised setting. ROC curves for both CK8/18 and H\&E based predictions for 5-fold cross-validation are shown in Figure \ref{fig:roc_msi}. 

\begin{table}[]
\centering
\begin{tabular}{ll|l|l|l|}
\cline{3-5}
                                             &             & \multicolumn{1}{c|}{Linear} & \multicolumn{1}{c|}{Polynomial} & \multicolumn{1}{c|}{RBF} \\ \hline
\multicolumn{1}{|l|}{\multirow{2}{*}{CK8/18}} & Localised   & 0.9596$\pm$0.048             & 0.9444$\pm$0.062                 & \textbf{0.9641$\pm$0.040}          \\ \cline{2-5} 
\multicolumn{1}{|l|}{}                       & Unlocalised & 0.9051$\pm$0.124             & 0.9414$\pm$0.055                 & \textbf{0.9429$\pm$0.067}          \\ \hline
\multicolumn{1}{|l|}{\multirow{2}{*}{H\&E}}  & Localised   & 0.8152$\pm$0.176             & 0.8455$\pm$0.149                 & \textbf{0.9217$\pm$0.115}        \\ \cline{2-5} 
\multicolumn{1}{|l|}{}                       & Unlocalised & 0.7798$\pm$0.179           & 0.7859$\pm$0.175                 & \textbf{0.8313$\pm$0.180}          \\ \hline
\end{tabular}
\caption{Average AUROC values for MSI prediction using 5-fold cross-validation obtained with different kernel functions in SVM. }
\label{tab:auc_ck818_he}
\end{table}

\subsection{Comparative Analysis}
MSI cases can have different patterns of MMR expression, as observed in our dataset as well (see Supplementary Table 1). Hence, in this regard, MMR labels can be considered as fine-grained labels of the coarse-grain MSS/MSI labelling. To the best of our knowledge, it is the first time that MMR labels have been utilised for microsatellite (MS) classification. It is equally correct assigning MSI status to a case having an MMR pattern where at least one of the MMR proteins has a loss status. Since our main task at hand in this study was to classify coarse-grained labels (MSS vs MSI), a question can arise: why not train a CNN using coarse-grain labels as it has been done in previous studies on MSI classification \cite{kather2019deep, kather2020pan, echle2020clinical, muti2021development, schmauch2020deep, bilal2021novel}. In other words, is it beneficial to first train a CNN model with fine-grain labels (MMR labels) and then map them to coarse-grain labels (MSI/MSS) in a second phase? 

Chen \textit{et al.} \cite{chen2018understanding} have shown that the CNN trained on natural images with fine-grained labels achieves better performance as compared to the network trained with coarse-grained labelled images of the same dataset. To study this with our dataset, we conducted another set of experiments, in which CNN was directly trained on MSI/MSS labelled patches instead of MMR labels. We trained the Xception network using Adam optimiser with binary cross-entropy as an objective function. The samples with a high proportion of MSI predicted patches to the total sum of MSI and MSS patches were labelled as MSI cases. We obtained AUC of $0.8646 \pm 0.1152$ and $0.8152 \pm 0.1019$ using the CK8/18 and H\&E stained slides, respectively. In addition, similar to our approach discussed in section \ref{MSI_prediction}, we trained an SVM using two normalised histograms of probabilities: one for MSS and the other for MSI and obtained AUC of $0.8465 \pm 0.1391$ and $0.8152 \pm 0.1166$ with CK8/18 and H\&E stained slides, respectively. These results demonstrated that MSI classification can be improved on training CNN with fine-grain labels (MMR labels). 

\subsection{TCGA Cohort}
To investigate the generalisability of our trained CNN model, we predicted MMR status in the patient cohorts from The Cancer Genome Atlas (TCGA). We experimented with the colon and rectal cancer cohort (TCGA-CRC-DX) which was used in previous studies \cite{kather2019deep, echle2020clinical, bilal2021novel} on MSI prediction. The TCGA-CRC-DX cohort comprises a total of 428 WSIs, obtained from 423 patients. The ground truth labels of the cohort contain only MSI labels, not MMR expression status. To this end, we retrained our multi-headed network with all the H\&E stained WSIs in the COMET dataset, followed by MMR prediction on the TCGA dataset. The training was performed in two separate settings: with stain augmentation and stain normalisation. 

A few example images in MSI and MSS subgroups, with the overlaid predicted MMR heatmaps, are shown in Supplementary Figures 6 to 9. On visualising MMR prediction heatmaps overlaid on the H\&E images, we mainly observed two patterns of MMR expression: a pattern where most of the tumour regions are predicted as intact (or positive) for all the MMR markers (see Supplementary Figures 7 and 8) and patterns where most of the tumour regions are predicted as intact for MSH2 and MSH6 markers and loss (or negative) for MLH1 and PMS2 markers (see Supplementary Figures 6 and 9). We found that the trained model didn't generalise well for the TCGA dataset. Unlike our observation for the COMET cases, the aforementioned patterns were observed in both MSI and MSS cases, hence MMR heatmaps could not be used to extract discriminatory features for the TCGA dataset. We observed similar findings with both models, irrespective of their training strategies (stain augmentation or stain normalisation).

In computational pathology, model generalisation is a major challenge due to the domain shift. It is particularly prevalent when the model is trained on a small dataset, obtained from a single site with less diversity in terms of sample preparation and image acquisition. The problem of generalisation in our experiments is due to the same reason, the domain shift. Below, we list some possible elements contributing to the domain shift:

\begin{itemize}
  \item Tissue samples for COMET were collected from a single site and all WSIs were acquired using Omnyx VL120 scanner while following the same acquisition protocols. The TCGA slides are collected from many sites across the USA. In addition to the multi-site aspect, the heterogeneity can also be characterised by the scanner modalities, manufacturers and acquisition protocols \cite{tcga_coad}.
  
  \item The COMET and TCGA slides differ in terms of the stage of cancer. COMET slides were taken from the cancerous region described as either stage 3 or 4 while TCGA samples were taken from stage 1-4 cancers. 
  
  \item We have looked at the distribution of histological type by MSI status in mucinous and non-mucinous cancer for both the COMET and TCGA cohort, as shown in Supplementary Figure 10. The number of mucinous samples with both MSI and MSS status is significantly low for both datasets. Since our CNN model is trained on the COMET dataset, a total of 4 and 2 mucinous samples for MSI and MSS respectively are not adequate for learning discriminatory features for classification.
  
  \item The genetic mutations are identified using different methods for the TCGA and COMET datasets. PCR test is used for the TCGA dataset whereas IHC is utilised for the COMET dataset. However, the rate of discordant results between IHC and PCR is very low (2\%) and is shown to be due to defective MMR \cite{discordant_IHC_PCR}.
\end{itemize}


\section{Conclusion}
In this paper, we have presented a novel method for MSI classification in a two-stage process. The main focus is to perform MMR biomarker estimation from a single target image (CK8/18 or H\&E) with an end goal of identifying cases with MSI status. To this end, we utilised registered images of the target and MMR tissue slides for identifying tissue regions with stain intact/loss status in a localised manner, rather than assigning a WSI label to every patch. Even with imperfections in MMR prediction maps, SVM was able to identify samples with MSI status. Our comparative analysis demonstrated improved performance of the CNN model if trained in a localised setting. Similar to the findings in previous literature, CNN trained with fine-grain MMR classes outperformed the CNN trained with coarse-grain MSI/MSS classes.

In clinical practice, identification of dMMR cases either using IHC or PCR testing incurs significant additional costs and requires additional resources which may not be available in all pathology laboratories. There is also the factor of the additional time required to perform these tests which can delay the initiation of appropriate treatment. As the adoption of digital pathology continues to grow and advance, the introduction of computational tools may assist in overcoming these issues and provide a fast and accurate way to identify and stratify patients with dMMR CRC so that they can be offered the most appropriate treatment pathways. The results of biomarker status prediction using computational methods such as those presented in our study are encouraging and can open routes for their applications in clinical workflows after further improvement and validation at a larger scale.



\bibliographystyle{vancouver}
\bibliography{sample}

\end{document}